\documentstyle[11pt,newpasp,twoside,epsf]{article}
\def\edcomment#1{\iffalse\marginpar{\raggedright\sl#1\/}\else\relax\fi}
\reversemarginpar

\begin{document}
\title{Diffuse Cluster-wide Radio Halos}
 \author{Haida Liang}
\affil{Physics Dept., University of Bristol, Tyndall Ave., Bristol BS8 1TL, UK}

\begin{abstract}
We will discuss the properties
and origins of halos and relics including estimates of the cluster
magnetic fields, and present results for a few recently discovered
halos and relics. The electrons in the
suprathermal high energy tail of the thermal gas distribution are
likely to provide the seed particles for acceleration through mergers
and turbulences to relativistic energies. These relativistic particles
are then responsible for the synchrotron emission of the halos.
\end{abstract}

\section{Properties of Radio Halos}
Radio halos are preferentially found in high X-ray luminosity clusters
(Giovannini et al. 1999) and a tentative correlation was found between
the radio luminosity of halos and X-ray temperature of cluster gas
(Liang et al. 2000).  Recently, we have observed a number of clusters
with relatively low temperature ($kT_{x}<7$\,keV) to test if this
correlation still holds and whether there is a switch-on temperature
for radio halo production. We obtained upper-limits from ATCA
observations of the clusters A3158, A3921, and updated the
$P_{1.4}-kT_{x}$ plot (Fig.~1) with the addition of A2199 (Kempner \&
Sarazin 2000) and A3667 (Ekers private comm.). We also attempted to
test the theory that merging is the essential ingredient in halo
production, since so far all the known radio halos are detected in
non-cooling flow, merging clusters (Giovannini et al. 1999). For this
reason, we chose to observe with the ATCA a strong cooling flow
cluster (i.e. no merging), RXJ1347-11, with a high temperature
(Fig.~1). We were not able to reach a low brightness limit because of
the relatively strong central radio source in the cluster. The cluster
was also fairly small in angular size which made it even more
difficult to separate the central radio source from a cluster-wide
diffuse emission. These features are common to most cooling flow
clusters, which may cause a bias in the result that only non-cooling
flows have radio halos. Note that we prefer the $P_{1.4} - kT_{x}$
relation to the $P_{1.4} - L_{x}$ relation because it is free from the
Malquist bias common to luminosity-luminosity relations; furthermore
$kT_{x}$ is directly related to the cluster total mass.

On the large scale, radio halos have similar extent and shape as the
thermal X-ray emission (e.g. Fig.~1). However, the radio emission
corresponds better in detail to the galaxy distribution than the X-ray
emission in the case of 1E0657-56 (Liang et al. 2000) and Coma (Kim et
al. 1990). In 1E0657-56 and A2744, the clusters have substructure in
both X-ray and galaxy distributions with the X-ray subclumps displaced
from the galaxy subclumps (Fig~1).

\section{Intracluster Magnetic field}
Under the assumption of equipartition, cluster $B$-fields are
estimated to be of order of a few $\mu$G from the radio brightness of
the cluster-wide halos.  The first direct measurement of a cluster
$B$-field was in the Coma cluster ($B\sim 2\mu$G), through
measurements of the Faraday rotation effects in polarised sources seen
through the cluster (Kim et al. 1990). Recently, Clarke et al. (1999)
showed, in their statistical analysis of the rotation measure (RM) of
polarised radio sources seen through 16 low redhsift ($z<0.1$)
clusters without radio halos, that the cluster magnetic fields are
$B\sim 5-10\mu$G. Another method of directly measuring the
cluster-wide $B$-field is from the detection of the inverse Compton
(IC) scattered hard X-ray emission in excess of the thermal
Bremsstrahlung emission.  Recently, excess hard X-ray emission has
been detected in the Coma cluster from Beppo-SAX, which gave a cluster
$B$-field of $\sim 0.2\mu$G (Fusco-Femiano et al. 1999). However,
there are still debates over the origin of the detected hard X-ray
excess, with suggestions of it being Bremsstrahlung radiation of
supra-thermal electrons (e.g. Dogiel 2000).
\begin{figure}
\plottwo{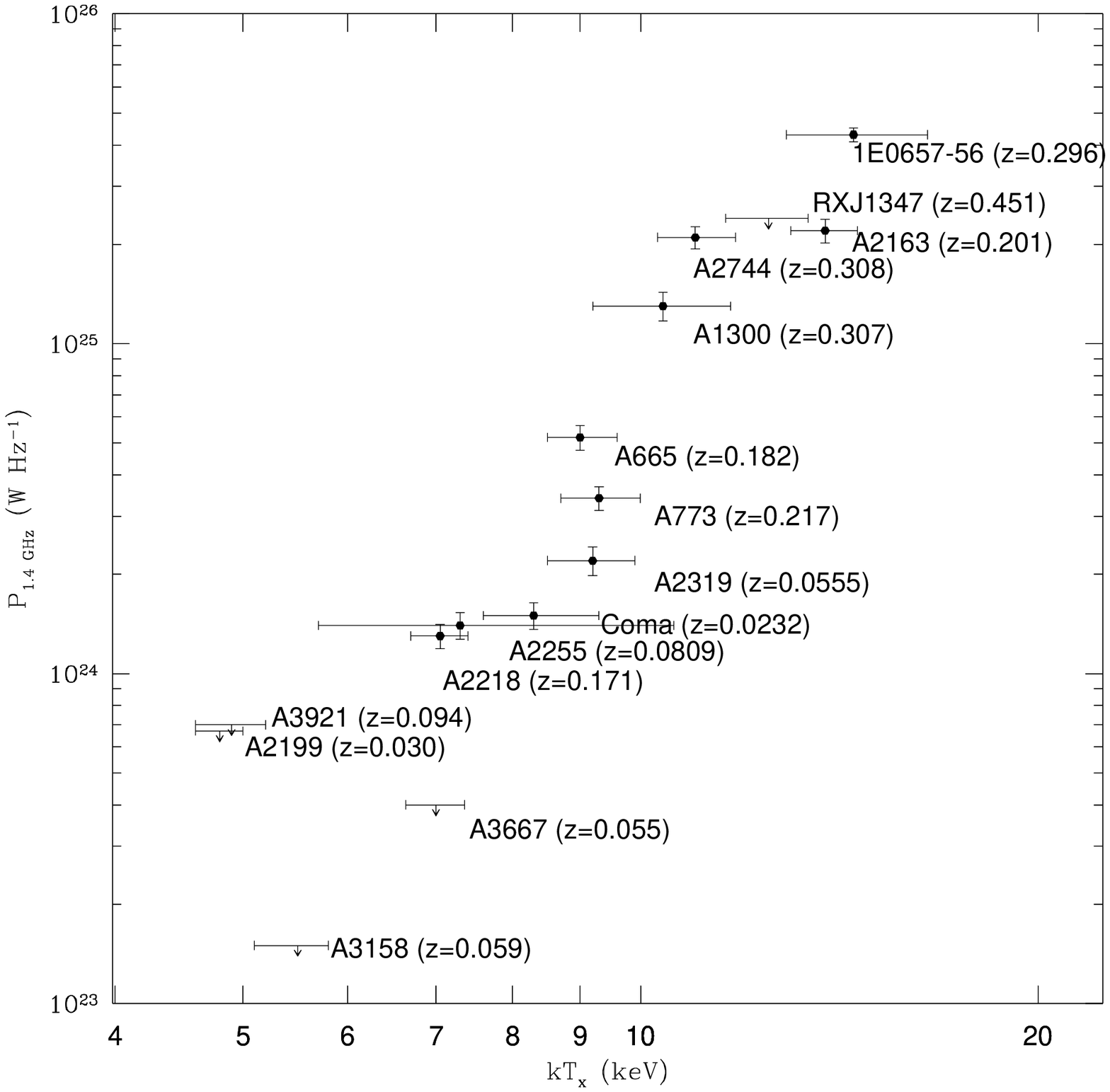}{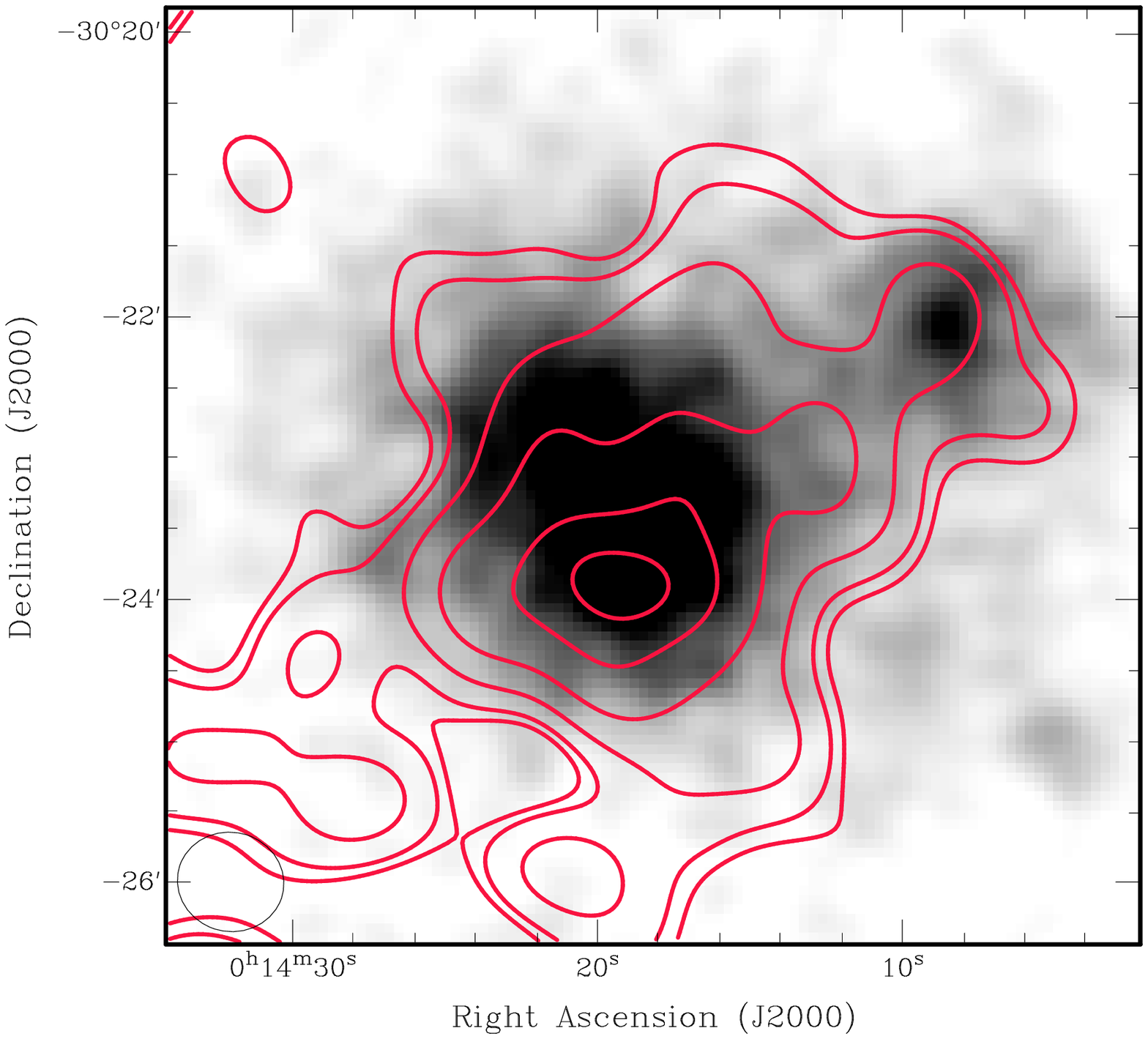}
\caption{LEFT: Radio power $P_{1.4}$ at 1.4 GHz (rest frame) versus
the cluster X-ray temperature $kT_{x}$. Similar to Fig.~9 in Liang et
al. (2000) but with the addition of a few upper limits. RIGHT: A grey
scale ROSAT PSPC image of A2744 overlaid with contours of 1.3\,GHz
ATCA image of the radio halo.}
\end{figure}

So far no polarisation has been detected in any known radio halo. In
the case of 1E0657-56, the upper limit to the percentage polarisation
is $<20$\% on 55\,kpc scales at 1.3\,GHz. The polarisation upper
limits at higher frequencies are in general poorer because of the
steepness of the halo spectral indices.  Radio relics, on the other
hand, are found to be polarised. In 1E0657-56, the relic source is
$\sim 10$\% polarised at 1.3\,GHz with mild depolarisation. Similarly,
in A2256 (R\"ottgering et al. 1997) and A3667 (Johnston-Hollitt
private comm.), depolarisation is only mild in the radio relics
suggesting that relics are likely to be at the front of clusters. We
deduce the B-field to be $>0.1\mu$G from the average rotation measures
of the relic.  The rotation measure deduced from an
extended, steep spectra, highly polarised radio source (54\% at
8.8\,GHz), J0658-5559, also gave a lower limit of $0.2\mu$G to the B-field of
1E0657-56 (Liang et al. in prep.).

\section{Model for Radio Halo Production}
The steep correlation $P_{1.4}-kT_{x}$ and the similarity in X-ray and
radio halo morphology suggest that the electrons in the suprathermal
high energy tail of the thermal gas distribution are likely to provide
the seed particles for acceleration through mergers and turbulences to
relativistic energies. These relativistic particles are then
responsible for the production of the synchrotron emission.  Dogiel
(2000) pointed out that the hard X-ray excess (30-80\,keV) observed in
Coma by Beppo-SAX can be the result of Bremsstrahlung radiation from
suprathermal electrons in the tail of the distorted Maxwellian
spectrum. If there is in-situ acceleration, then the particle spectrum
starts to deviate from the Maxwellian even below the injection energy
due to Coulomb collisions.  In this model, the only free parameters
are the acceleration parameter and the particle production spectral
index. Here we examine if the non-thermal tail of the Maxwellian can
produce sufficient radio emission to explain the halo in Coma. The
spectral index of the halo fixes the particle production spectral
index and consequently the acceleration parameter given the observed
hard X-ray excess. A magnetic field of $B\sim 1.5\mu$G is necessary to
produce the observed radio halo flux density (Dogiel \& Liang in
prep.). In comparison, the magnetic field deduced from Faraday
rotation gives $B<6h_{50}^{1/2}\mu$G on scales $<1$\,kpc with a large
scale (200\,kpc) average magnetic field of $B\sim 0.2\mu$G (Feretti et
al 1995); and the equipartition B-field is $\sim
0.9\mu$G. Hence, the model successfully fits the hard X-ray excess and
the radio halo emission for the Coma cluster with a value of the
cluster magnetic field consistent with current observations. Note that
in this model, the IC radiation in the 30-80\,keV range is orders of
magnitude smaller than the Bremsstrahlung radiation.

\acknowledgements 
I thank my collaborators V. Dogiel, R. Hunstead, 
R. Ekers, M. Birkinshaw, P. Andreani, P. Shaver, I. Hook, W. Couch \& E. Falco
for valuable contributions; and Prof. R. Buonanno for hospitality at the OAR.

\end{document}